\documentclass[]{elsarticle}

\usepackage{lineno,hyperref}

\journal{Journal of \LaTeX\ Templates}

\newtheorem{dfn}{Definition}
\newtheorem{cor}{Corollary}
\newtheorem{prop}{Proposition}
\newtheorem{theorem}{Theorem}
\begin{document}

\begin{frontmatter}

\title{On a Relation among Bi-orthogonal system, Quadratic Non-Hermitian Boson operators with real spectrum and Partial $\mathcal{PT}$ symmetry in Fock Space}

\author{Arindam Chakraborty\fnref{myfootnote}}
\address{Department of Physics, Heritage Institute of Technology, Kolkata-700107, India}

\ead[url]{www.elsevier.com}

\cortext[mycorrespondingauthor]{Arindam Chakraborty}
\ead{arindam.chakraborty@heritageit.edu}


\begin{abstract}
A new kind of symmetry called partial $\mathcal{PT}$ symmetry (henceforth $\partial_{\mathcal{PT}}$) has been considered for non-hermitian quadratic boson operators obtained from a bi-orthogonal set of vectors in $\mathbf{C}^2$. The symmetry behaviour has been understood in Fock space considered as a Reproducing Kernel Hilbert Space(RKHS). The reality of eigenvalues and its connection to the possibility of the aforesaid symmetry  (and symmetry breaking) are studied in terms of a deformation parameter responsible for non-hermiticity.
\end{abstract}


\end{frontmatter}


\section{Introduction}

In recent years the use of non-hermitian operators with real spectrum in the context of space-time reflection symmetry (${\mathcal {PT}}$-symmetry) has gained much relevance in modern quantum methods  \cite{bender98, bender99, bender02,mosta02,roy06, brody14,yuce15, brody16,}. The issue of ${\mathcal {PT}}$-symmetry in non-hermitian Hamiltonian \cite{garcia17} has been widely studied  in open regimes with balanced gain and loss like laser absorbers \cite{longhi10}, ultra-cold threshold phonon lasers \cite{jing14}, defect states in special beam dynamics in spatial lattices \cite{rogen13},quantum circuit based on nuclear magnetic resonance \cite{zheng13}, microwave cavities \cite{bitt12}, super-conductivity \cite{rubin07}, Bose-Einstein condensates \cite{graefe08, krei16}, optical wave guides \cite{klai08} and quantum entanglement\cite{gopal21}. The theoretical developments in terms of space and algebraic structures of non-hermitian systems \cite{bagchi00, bagchi02,brody17}, the role of ${\mathcal {PT}}$-symmetric operators  and their significance in pseudo-hermitian quantum mechanics \cite{mosta05} have also become some interesting areas of investigation \cite{baga15, moise11, bender19}.

The extension of symmetry argument for systems involving two or more particles  has been attempted earlier\cite{graefe08, nana02}. However, the reality of the spectrum does not necessarily correspond to the existence of global ${\mathcal {PT}}$-symmetry. In fact in  a recent article by Beygi et. al. \cite{beygi15}  a variable dependent $\mathcal{PT}$ symmetry behaviour has been investigated in a model of coupled oscillator with purely imaginary coupling term along with the possibilities of real and imaginary eigenvalues. Our present objective is to elaborate the symmetry argument in a \textbf{Fock Space} setting with the help of \textbf{weighted composition conjugation} \cite{hai18,hai16,hai18c} representing complex symmetries \cite{gar05,gar07} and acting on a \textbf{Reproducing Kernel Hilbert Space}\cite{raghu16, sai16}. The following discussion also shows that a \textbf{spectrum generating algebra} can be constructed starting from a \textbf{bi-orthogonal} system of vectors in $\mathbf{C}^2$ and reality of the spectrum has a direct relation with bi-orthogonality as well as with the existence and breaking of $\partial_{\mathcal{PT}}$-symmetry.

\section{Quadratic Boson Operator and $\partial_{\mathcal{PT}}$ symmetry}
We consider a specific quadratic boson operator $B=\frac{1}{2}\sum_{m,n=1}^3B_{mn}a^{\dagger}_{m}a_{n}$ where, $\{a_\alpha, a^\dagger_\alpha \vert\alpha = 1, 2\}$ are boson operators with $[a_\alpha, a^\dagger_\beta] - \delta_{\alpha\beta}=[a_{\alpha}, a_{\beta}]=[a^{\dagger}_{\alpha}, a^{\dagger}_{\beta}]=0$ and $B_{11}=c_1, B_{22}=c_2, B_{12}=B_{21}=i\alpha$, $c_1$, $c_2$ and $\alpha$ being real.
We consider  $(\mathcal{F}^2(\mathbf{C}^2))$ as the  Fock (or Segal-Bargmann) space which is a separable complex Hilbert space of entire functions (of the complex variables $\zeta_1$ and $\zeta_2$) equipped with an inner-product 
\begin{eqnarray*}
\langle \psi, \phi\rangle=\int_{\mu(\zeta_1)}\int_{\mu(\zeta_2)} \psi(\zeta_1,\zeta_2)\overline{\phi(\zeta_1, \zeta_2)}\\
\;\; {\rm{with}} \int_{\mu(\zeta_1)}\int_{\mu(\zeta_2)}\equiv \int\int d{\mu(\zeta_1)}d{\mu(\zeta_2)}\nonumber\\
\end{eqnarray*}
Here, ${d}{\mu(z)}=\frac{1}{\pi}e^{-\vert z\vert^2}{d}({\rm{Re}}(z)) d({\rm{Im}}(z))$ represents the relevant Gaussian measure relative to the complex variable $z$.

\begin{dfn}
In a Fock space of one complex variable $\zeta$, the \textbf{weighted composition conjugation} \cite{hai18} can be given by the expression
\begin{eqnarray}
\mathcal{W}_{(\vartheta, \eta, \upsilon)}\psi(\zeta)=\upsilon e^{\eta\zeta}\overline{\psi(\overline{\vartheta\zeta+\eta})}.
\end{eqnarray}
\end{dfn}

Here, $\zeta$ is a complex variable and $\{\vartheta, \eta, \upsilon\}$ are complex numbers satisfying the set of necessary and sufficient conditions : $\vert\vartheta\vert=1,\bar{\vartheta}\eta+\bar{\eta}=0$ and $\vert\upsilon\vert^2e^{\vert\eta\vert^2}=1$. The anti-linear operator $\mathcal{W}_{(\vartheta, \eta, \upsilon)}$ is  involutive and isometric and hence a \textbf{conjugation}. The action of the operator $\mathcal{PT}$ is equivalent to the choice : $\vartheta=-1=-\upsilon,\eta=0$ which gives
$\mathcal{W}_{(\vartheta, 0, 1)}\vert_{\vartheta=-1}\psi(\zeta)=\overline{\psi(\overline{-\zeta})}.$
Similarly, the action of $\mathcal{T}$ involves the choice : $\vartheta=1, \eta=0, \upsilon=1$ giving
$\mathcal{W}_{(\vartheta, 0, 1)}\vert_{\vartheta=1}\psi(\zeta)=\overline{\psi(\overline{\zeta})}$.

Considering $\psi\dot{=}\psi(\zeta_1,\zeta_2,\cdots \zeta_n)$ as a function of several complex variables  one can define an operator $\mathcal{W}_{(\vartheta_j,\eta_j, \upsilon_j : j=1\cdots n )}$ with the action
\begin{eqnarray}
\mathcal{W}_{(\vartheta_j,\eta_j= 0,\upsilon_j= 1 : j=1\cdots n)}\psi(\zeta_1,\cdots,\zeta_j,\cdots,\zeta_n)=\overline{\psi(\overline{\vartheta_1\zeta_1},\cdots,\overline{\vartheta_j\zeta_j},\cdots, \overline{\vartheta_n\zeta_n})}.
\end{eqnarray}
\begin{dfn} 
An operator $\mathcal{W}^{(j)}_n=\mathcal{W}_{(\vartheta_j,\eta_j= 0,\upsilon_j= 1 ; j=1\cdots n)}\vert_{\vartheta_1=1,\cdots,\vartheta_j=-1,\cdots,\vartheta_n=1}$ is called the \textbf{$j$-th partial $\mathcal{PT}$ symmetry ($\partial_{\mathcal{PT}}$) operator} if
\begin{eqnarray}
\mathcal{W}^{(j)}_n\psi(\zeta_1,\cdots,\zeta_j,\cdots,\zeta_n)=\overline{\psi(\bar{\zeta_1},\cdots,\overline{-\zeta_j},\cdots,\bar{\zeta_n})}.
\end{eqnarray}
\end{dfn}
and an  operator $\mathcal{W}_n$ is called a \textbf{global $\mathcal{PT}$ symmetry operator} if
\begin{eqnarray}
\mathcal{W}_n\psi(\zeta_1,\cdots,\zeta_j,\cdots,\zeta_n)=\overline{\psi(\overline{-\zeta_1},\cdots,\overline{-\zeta_j},\cdots,\overline{-\zeta_n})}.
\end{eqnarray}
For our present purpose we shall only consider the operators $\mathcal{W}_2$ and $\{\mathcal{W}_2^{(j)} : j=1,2\}$. Now, global and partial  $\mathcal{PT}$ symmetries of any
function $\psi(\zeta_1, \zeta_2)$ are expressed through the relations
$\mathcal{W}_2\psi(\zeta_1, \zeta_2)=\psi(\zeta_1, \zeta_2)$ and $\mathcal{W}_2^{(j)}\psi(\zeta_1, \zeta_2)=\psi(\zeta_1, \zeta_2)\:\:\forall\:\: j=1, 2$
respectively.

For the symmetry behaviour of $B(c_1, c_2, i\alpha, i\alpha)$ we consider the following notion of \textbf{Reproducing Kernel Hilbert Space (RKHS)}.
\begin{dfn}
A function of the form $\kappa^{[m_1, m_2]}_{\zeta_1, \zeta_2}(z_1, z_2)=z_1^{m_1}z_2^{m_2}e^{z_1\overline{\zeta_1}+z_2\overline{\zeta_2}}$ with $m_1, m_2\in\mathbf{N}$ and $\zeta_j,z_j\in \mathbf{C}\:\:\forall\:\:j=1,2$ is called a kernel function (or a \textbf{Reproducing Kernel}) which satisfies the condition 
\begin{eqnarray}
	\psi^{(m_1,m_2)}(\zeta_1, \zeta_2)=\left\langle\psi, \kappa^{[m_1, m_2]}_{\zeta_1, \zeta_2}\right\rangle=\int_{\mu(z_1)}\int_{\mu(z_2)}\psi(z_1, z_2)\overline{\kappa^{[m_1, m_2]}_{\zeta_1, \zeta_2}}. 
	\end{eqnarray}
\end{dfn}
Here, $\psi\in\mathcal{F}^2(\mathbf{C}^2)$, $\psi^{(m_1, m_2)}(\zeta_1, \zeta_2)=\partial_{\zeta_1}^{m_1}\partial_{\zeta_2}^{m_2}\psi$ and $\psi^{(0, 0)}(\zeta_1, \zeta_2)\equiv\psi(\zeta_1, \zeta_2)$. Hence the
\begin{prop}
	$B(c_1, c_2, i\alpha, i\alpha)^{\star}\neq B(c_1, c_2, i\alpha, i\alpha)$ where for any operator $H$, $H^{\star}$ is defined as $\left\langle H\psi(u_1,u_2), \kappa^{[m_1, m_2]}_{\zeta_1, \zeta_2}\right\rangle=\left\langle\psi(u_1,u_2), H^{\star}\kappa^{[m_1, m_2]}_{\zeta_1, \zeta_2}\right\rangle$. 
\end{prop}
The self-adjointness of the operator $B$ we can write the 
\begin{prop}
	$B(c_1, c_2, i\alpha, i\alpha)$ is $\mathcal{W}_2$-self-adjoint i. e.; $\mathcal{W}_2B(c_1, c_2, i\alpha, i\alpha)^{\star}\mathcal{W}_2=B(c_1, c_2, i\alpha, i\alpha)$ but it is not $\mathcal{W}_2^{(j)}$-self-adjoint.		
\end{prop}

Proof :
	We shall first show the case with $z_1\partial_{z_1}$. Considering the conjugation operator $\mathcal{W}_{2}$ we find
	\begin{eqnarray}
	&&\mathcal{W}_{2}(z_1\partial_{z_1})^{\star}\mathcal{W}_{2}\kappa^{[m_1, m_2]}_{\zeta_1, \zeta_2}(z_1, z_2)\nonumber\\
	&=&\mathcal{W}_{2}z_1\partial_{z_1}\mathcal{W}_{2}z_1^{m_1}z_2^{m_2}e^{z_1\overline{\zeta_1}+z_2\overline{\zeta_2}}\nonumber\\
	&=&\mathcal{W}_{2}z_1\partial_{z_1}\overline{\overline{(-z_1)}^{m_1}}\overline{\overline{(-z_2)}^{m_2}}\overline{e^{\overline{-z_1}\overline{\zeta_1}+\overline{-z_2}\overline{\zeta_2}}}\nonumber\\
	&=&\mathcal{W}_{2}z_1\partial_{z_1}(-1)^{m_1+m_2}z_1^{m_1}z_2^{m_2}e^{-z_1{\zeta_1}-z_2{\zeta_2}}\nonumber\\
	&=&\mathcal{W}_{2}z_1(-1)^{m_1+m_2}[z_1^{m_1}z_2^{m_2}(-{\zeta_1})e^{-z_1{\zeta_1}-z_2{\zeta_2}}+m_1z_1^{m_1-1}z_2^{m_2}e^{-z_1{\zeta_1}-z_2{\zeta_2}}]\nonumber\\
	&=&-z_1(-1)^{m_1+m_2}[(-1)^{m_1+m_2}z_1^{m_1}z_2^{m_2}(-\overline{\zeta_1})\nonumber\\
	&&+m_1(-1)^{m_1+m_2-1}z_1^{m_1-1}z_2^{m_2}]e^{z_1\overline{\zeta_1}+z_2\overline{\zeta_2}}\nonumber\\
	&=&z_1\partial_{z_1}\kappa^{[m_1, m_2]}_{\zeta_1, \zeta_2}(z_1, z_2).
	\end{eqnarray}
Similarly one can prove \[\mathcal{W}_{2}(iz_1\partial_{z_2})^{\star}\mathcal{W}_{2}\kappa^{[m_1, m_2]}_{\zeta_1, \zeta_2}(z_1, z_2)=(iz_2\partial_{z_1})\kappa^{[m_1, m_2]}_{\zeta_1, \zeta_2}(z_1, z_2)\] and \[\mathcal{W}_{2}(iz_2\partial_{z_1})^{\star}\mathcal{W}_{2}\kappa^{[m_1, m_2]}_{\zeta_1, \zeta_2}(z_1, z_2)=(iz_1\partial_{z_2})\kappa^{[m_1, m_2]}_{\zeta_1, \zeta_2}(z_1, z_2).\] Using these results in the expression of the $B(c_1, c_2, i\alpha, i\alpha)$, the above proposition is verified. Similar calculation may be repeated to show the absence of $\mathcal{W}_2^{(j)}$-self-adjointness. 

Similar argument justifies the following proposition
 \begin{prop}
	$B(c_1, c_2, i\alpha, i\alpha)$ has $\partial_{\mathcal{PT}}$ symmetry i. e.; $\mathcal{W}_2^{(j)}B(c_1, c_2, i\alpha, i\alpha)$ $\mathcal{W}_2^{(j)}=B(c_1, c_2, i\alpha, i\alpha)$, for $j=1, 2$ but it lacks global $\mathcal{PT}$ symmetry i. e.; 
	$\mathcal{W}_2B(c_1, c_2, i\alpha, i\alpha)$ $\mathcal{W}_2\neq B(c_1, c_2, i\alpha, i\alpha)$.  
\end{prop}	

\section{$\partial_{\mathcal{PT}}$ symmetry of the eigenstates of $B(c_1, c_2, i\alpha, i\alpha)$ }
First we shall show that the relation between the reality of the eigenvalues and $\partial_{\mathcal{PT}}$ symmetry of the eigen states of the Hamiltonian. Since that the present operator leaves the homogeneous polynomial space of two indeterminates $(\zeta_1, \zeta_2)$ invariant, let us consider the degree of homogeneity $m$ and polynomial bases $\{f_{k} = \zeta_1^{m-k} \zeta_2^{k}
: k = 0 \cdots m\}$. The operator $B(c_1, c_2, i\alpha, i\alpha)$  takes the following tridiagonal representation

\begin{eqnarray*}\label{tri1}
B
=&& 
\frac{1}{2}\left( \begin{array}{cccccccc}
\beta_{m.0} & i\alpha & 0 & 0 & 0& \cdots  & 0& 0 \\
im\alpha & \beta_{m,1} & 2i\alpha & 0 & 0 &\cdots &0 & 0 \\
0& i\alpha(m-1) & \beta_{m, 2} & 3i\alpha & 0 & \cdots &0 & 0 \\
\vdots & \vdots & \vdots & \vdots & \vdots & \cdots & \beta_{m, m-1} & im\alpha \\
0 & 0& 0& 0& 0& \cdots & i\alpha & \beta_{m, m}
\end{array} \right).  
\end{eqnarray*}
where,  $\{\beta_{m,k}=c_1(m-k)+c_2k : k=0,\dots, m\}$
The the eigenvalues of such a matrix can be found out with the help of the following theorem \cite{sandry13}.
\begin{theorem}
	Given a tri-diagonal matrix
	\begin{equation}\label{tri2}
	{\mathcal M} = \left(
	\begin{array}{cccccccc}
	b_0 & d_0 & 0 & 0 & 0 & \cdots & 0 & 0 \\
	c_0 & b_1 & d_1 & 0 & 0 & \cdots & 0 & 0 \\
	0 & c_1 & b_2 & d_2 & 0 & \cdots & 0 & 0 \\
	\vdots & \vdots & \vdots & \vdots & \vdots & \cdots & b_{l-2} & d_{l-2} \\
	0 & 0 & 0 & 0 & 0 & \cdots & c_{l-2} & b_{l-1}
	\end{array}\right)
	\end{equation}
	with $d_i \neq 0 \:\:\forall\:\: i$, let us consider a polynomial $Q_n (\lambda)$
	that follows the well-known three term recursion relation \cite{dunkl14, ismail05} 
	\begin{equation}\label{rec1}
	Q_{n+1}(\lambda) = \frac{1}{d_n}[(\lambda - b_n )Q_n (x) - c_{n-1} Q_{n-1} (\lambda)].    
	\end{equation}
If $Q_{-1}(\lambda) = 0$ and $Q_0(\lambda) = 1$ the eigenvalues are given by the zeros
	of the polynomial $Q_l(\lambda)$ and eigenvector corresponding to $j$-th eigenvalue
	$\lambda_j$ is given by the vector
	$\left(Q_0 (\lambda_j ) \:\:
	Q_1 (\lambda_j )\:\:
	\dots \:\:
	Q_{l-2} (\lambda_j)\:\:
	Q_{l-1} (\lambda_j ) \right)^{\dagger}$
\end{theorem}
Concerning the symmetry of the eigenfunctions follows the 
\begin{cor}
	If an eigenvalue of $B(c_1, c_2, i\alpha, i\alpha)$ is real the corresponding eigenfunction in the homogeneous polynomial space is $\partial_{\mathcal{PT}}$ symmetric (or anti-symmetric). If an eigenvalue has non-zero imaginary part the said symmetry is broken. 
\end{cor}
Proof : For real values of $c_1, c_2$ and $\alpha$ the set of polynomials ${Q_n(\lambda)}$ becomes alternatively real and purely imaginary for any real value of $\lambda$. The eigenfunction corresponding to $\lambda$ in any polynomial space with degree of homogeneity $m$ can be given by
$\psi_{2s-1}(\zeta_1, \zeta_2)=A_0\zeta_1^{2s-1}+iA_1\zeta_1^{2s-2}\zeta_2+A_2\zeta_1^{2s-3}\zeta_2^2+\dots+iA_{2s-1}\zeta_2^{2s-1}\:\:{\rm for}\:\: m=2s-1$ and $\psi_{2s}(\zeta_1, \zeta_2)=B_0\zeta_1^{2s}+iB_1\zeta_1^{2s-1}\zeta_2+B_2\zeta_1^{2s-2}\zeta_2^2+\dots+B_{2s}\zeta_2^{2s}
\:\:{\rm for}\:\: m=2s$. Here, $\{A_l\}$ and $\{B_l\}$ are real functions for real values of $\lambda$ with $A_0=B_0=1$. Now the action of $\mathcal{W}_2^{(j)}$ on $\psi_{2s-1}$ is given by
\begin{eqnarray}
	\mathcal{W}_2^{(1)}\psi_{2s-1}&=&\overline{A_0\overline{(-\zeta_1)}^{2s-1}+iA_1\overline{(-\zeta_1)}^{2s-2}\overline{(\zeta_1)}
	+\dots+A_{2s-1}\overline{(\zeta_2)}^{2s-1}}\nonumber\\
=-\psi_{2s-1}
\end{eqnarray}
Similarly, $\mathcal{W}_2^{(2)}\psi_{2s-1}=\psi_{2s-1}$ and $\mathcal{W}_2^{(j)}\psi_{2s}=\psi_{2s}\:\:\forall\:\: j=1,2$.
It can also be verified in the same manner that $\mathcal{W}_2^{(j)}\psi_{2s-1}\neq\pm\psi_{2s-1}$ and $\mathcal{W}_2^{(j)}\psi_{2s}\neq\psi_{2s}$ $\:\:\forall\:\: j=1,2$ if the eigenvalue has any non-zero imaginary part.

\section{A special case of $B(c_1,c_2,i\alpha,i\alpha)$  related to biorhogonal system and a deformed su(2) algebra}
\begin{dfn}
	Two pairs of vectors $\{\vert\phi_j\rangle : j=1,2\}$ and $\{\vert\chi_j\rangle : j=1,2\}$ are called bi-orthogonal if $\langle\phi_j\vert\chi_k\rangle=0\:\:\forall\:\: j\neq k$\cite{baga15, hajek08}.
\end{dfn}
Considering a pair of orthogonal vectors 
$\{\vert u_j\rangle = \frac{1}{\sqrt{2}}
\left(
	1 \:\:
	(-1)^{j-1}
\right)^{\dagger}: j = 1,2\}$, $\langle u_j\vert u_k\rangle = \delta_{jk}$ we get the set of bi-orthogonal vectors
$\vert\phi_j\rangle=\omega T\vert u_j\rangle$ and $\vert\chi_j\rangle=(T^{\dagger})^{-1}\vert u_j\rangle$ for $\omega=\cos\theta$ and $j=1,2$. Here, $T=\cos(\frac{\theta}{2})\mathbf{1}_2-2\sin(\frac{\theta}{2})\sigma_2$ and $\omega=\sqrt{1-\alpha^2}$. This leads to a set of  deformed $su(2)$ generators
\begin{eqnarray}
	\sigma_m^{\alpha} = \frac{i^{m+1}}{2}\sum_{j, k=1}^2\frac{c_{jk}^{(m)}}{\omega^{\delta_{m2}}}\vert\phi_j\rangle \langle\chi_k \vert : m = 1, 2, 3
\end{eqnarray}
Here, $c_{jk}^{(1)} = (-1)^j \delta_{jk}$, $c_{jk}^{(3)} = (-1)^j c_{jk}^{(2)} = (1 -(-1)^j c_{jk}^{(1)})$.The generators $\{\sigma_m^\alpha
: m = 1, 3\}$ are non-hermitian in the conventional sense of inner product in $\bf{C^2}$. The corresponding Jordan-Schwinger map  is given by
\begin{eqnarray}
	J_m^{\alpha} = \frac{i^{m-1}}{2}\sum_{j, k=1}^2[c_{jk}^{(4-m)} + i^m (1 - \delta_{m2} )\alpha c_{jk}^{(m)}] a^\dagger_j a_k : m = 1, 2, 3.
\end{eqnarray}
The components of $J_m^{\alpha}$ have the following identifications : $J_1^{\alpha}=B(-i\alpha, i\alpha, 1, 1)$, $J_2^{\alpha}=B(0, 0, -i, i)$ and $J_3^{\alpha}=B(1, -1, i\alpha, i\alpha)$.  Introducing a new set of deformed ladder operators with exponent $p$, $p\in \mathbf{R}$ (set of real numbers),
$J_\pm^\alpha = \sum_{r=0}^1(\pm i)^r \omega^{r-p} J^\alpha_{r+1}$ and considering $J_3^\alpha = J_0^\alpha$
the following spectrum generating algebra becomes possible

\begin{equation}
[J_0^{\alpha}, J_\pm^{\alpha}] = \pm\omega(\alpha) J_\pm^\alpha
\end{equation}
\begin{equation}\label{deformed1}
	[J_+^\alpha , J_-^\alpha] = 2\omega^{-2p+1}(\alpha)J_0^\alpha.  
\end{equation}

It is to be noted that $J_0^{\alpha}$ is no longer hermitian and $(J^{\alpha}_+)^{\dagger}\neq J^{\alpha}_-$. The Killing metric tensor is given by
\begin{equation}
	(g_{jk}) = 2 \left( \begin{array}{ccc}
		\omega^2 & 0 & 0 \\
		0& 0 & 2\omega^{-2p+2}\\
		0 & 2\omega^{-2p+2} & 0  
	\end{array} \right). 
\end{equation}
As the Killing form is non-degenerate the algebra is semisimple. 
Casimir of this algebra ${\mathcal {C}_J^\alpha} = \omega^{-2} J_0^\alpha (J_0^\alpha \pm \omega ) + \omega^{2p-2} J^\alpha_{\mp} J_\pm^\alpha$.

\subsection{Spectrum and eigenfunctions of $J_0^{\alpha}$}
Writing $J_0^{\alpha}=\frac{1}{2}\mathcal{A}$ where, the entries of $\mathcal{A}$ are equated with $\mathcal{M}$ with entries $\{b_0\cdots b_{l-1}\}=\{m,\cdots -m\}$, $\{c_0\cdots c_{l-2}\}=\{im\alpha,\cdots i\alpha\}$ and $\{d_0\cdots d_{l-2}\}=\{i\alpha,\cdots im\alpha\}$,  
the spectrum of $\mathcal{A}$  is given by
\begin{eqnarray}
	\Lambda^{\omega}_{odd}&=&\{\lambda\}_{odd}\nonumber\\
	&=&\{-(2s-1)\omega, -(2s-3)\omega, \dots ,(2s-3)\omega, (2s-1)\omega\}
\end{eqnarray}
 for $m=2s-1.$ 
and
\begin{eqnarray}
	\Lambda^{\omega}_{even}&=&\{\lambda\}_{even}=\{-(2s)\omega, -(2s-2)\omega, \dots ,(2s-2)\omega, (2s)\omega\}
\end{eqnarray}
 for $m=2s.$ 
 
It is to be mentioned that \textbf{the reality of the eigenvalues depends on the parameter $\alpha$ and the condition of bi-orthogonality puts the restriction $\vert\alpha\vert<1$ which fixes the  eigenvalues on the real line}.  For $\alpha=1$ all the eigenvalues coalesce to zero which becomes an \textbf{exceptional point} (for which the geometric multiplicity of an eigenvalue is greater than its algebraic multiplicity). If $\alpha$ is allowed to be treated as a free real parameter there may occur a phase transition of eigenvalues from real to imaginary values when $\alpha$ becomes greater than one.
The eigenfunctions can be obtained in view of section-3. It can be easily verified that $\partial_{\mathcal{PT}}$ symmetry is completely lost for $\alpha>1$ as expected. In the following we shall consider the cases with $m=2$ and $m=3$.

 The eigenfunctions for $m=2$ corresponding to eigenvalues (of $\mathcal{A}$)  $-2\omega, 0, 2\omega$  can be given by
\begin{eqnarray*}
	\Psi_{-2}(\zeta_1, \zeta_2)&=&\zeta_1^2+2iG_+\zeta_1\zeta_2-G_+^2\zeta_2^2\nonumber\\
	\Psi_0(\zeta_1, \zeta_2)&=&\zeta_1^2+iG_0\zeta_1\zeta_2-\zeta_2^2\nonumber\\
	\Psi_2(\zeta_1, \zeta_2)&=&\zeta_1^2+2iG_-\zeta_1\zeta_2-G_-^2\zeta_2^2.
	\end{eqnarray*}
Here, $G_{\pm}=\left(\frac{1\mp\omega}{1\pm\omega}\right)^{\frac{1}{2}}$ and $G_0=\frac{2}{\sqrt{1-\omega^2}}$.

Similarly, for $m=3$ the eigenfunctions corresponding to the eigenvalues $\{-3\omega,-\omega, \omega, 3\omega\}$ are givan by
\begin{eqnarray*}
	\Phi_{-3} &=&\zeta_1^3+3iG_+\zeta_1^2\zeta_2-3G_+^2\zeta_1\zeta_2^2-G_+^3\zeta_2^3\nonumber\\
	\Phi_{-1} &=&\zeta_1^3+i(G_0+G_+)\zeta_1^2\zeta_2+(G_+^2-G_0^2(1-\omega_0))\zeta_1\zeta_2^2-iG_+\zeta_2^3\nonumber\\
	\Phi_{1} &=&\zeta_1^3+i(G_0+G_-)\zeta_1^2\zeta_2+(G_-^2-G_0^2(1+\omega_0))\zeta_1\zeta_2^2-iG_-\zeta_2^3\nonumber\\
	\Phi_{3} &=&\zeta_1^3+3iG_-\zeta_1^2\zeta_2-3G_-^2\zeta_1\zeta_2^2-G_-^3\zeta_2^3
\end{eqnarray*} 
$\partial_{\mathcal{PT}}$ symmetry of the above eigenfunctions and possibility of symmetry breaking can be understood in view of the discussion in section-3. The wave functions are non-trivially modified in the sense that letting $\gamma=0$ those corresponding to the hermitian case cannot be retrieved. 

\section{Conclusion}
The above discussion makes it evident that (i) the construction of the deformed algebra from bi-orthogonal vectors, (ii) reality of the eigenvalues of $J_0^{\alpha}$ and (iii) existence and breaking of $\partial_{\mathcal{PT}}$ symmetry of the eigenfunctions are interrelated. Construction of a suitable symmetry induced inner product space may provide further insights of such systems. Furthermore, the use of reproducing kernel method is not much in vogue  in conventional quantum methods. Investigation may be furthered along this direction as well.

\section{Acknowledgement} AC wishes to thank Dr. Baisakhi Mal for extending help in preparing the document.

\end{document}